# Low power continuous-wave nonlinear optics in silica glass integrated waveguide structures

**M. Ferrera, L. Razzari, D. Duchesne and R. Morandotti**
*INRS-EMT, 1650 Boulevard Lionel Boulet, Varennes, Québec, Canada, J3X 1S2*
[1]*Also with: Dipartimento di Elettronica, Università di Pavia, via Ferrata 1, 27100 Pavia, Italy*
**Z. Yang, M. Liscidini and J.E. Sipe**
*Department of Physics and Institute for Optical Sciences, University of Toronto, 60 St. George St. Toronto M5S 1A7 Ontario, Canada*
**S. Chu and B.E. Little**
*Infinera Corp, 9020 Junction Dr, Annapolis, Maryland, USA 94089*
**D. J. Moss**
*CUDOS, School of Physics, University of Sydney, New South Wales 2006 Australia*
*dmoss@physics.usyd.edu.au*

**Photonic integrated circuits (PICs) are a key component [1] for future telecommunication networks, where demands for greater bandwidth, network flexibility, low energy consumption and cost must all be met. The quest for all-optical components has naturally targeted materials with extremely large nonlinearity, including chalcogenide glasses (ChG) [2] and semiconductors, such as silicon [3] and AlGaAs [4]. Yet issues such as immature fabrication technologies for ChG, and high linear and nonlinear losses for semiconductors, motivate the search for other materials. Here we present the first demonstration of nonlinear optics in integrated silica based glass waveguides using continuous wave (CW) light. We demonstrate four wave mixing (FWM), with low (7mW) CW pump power at $\lambda$=1550nm, in high index doped silica glass ring resonators capable of performing in photonic telecommunications networks as linear filters [5]. The high reliability, design flexibility, and manufacturability of our device raises the possibility of a new platform for future low-cost nonlinear all-optical PICs.**

All-optical signal processing has been demonstrated in a variety of materials and devices. Examples include all-optical regeneration via FWM or cross and self phase modulation (XPM, SPM) in silicon nanowire waveguides [3,6] at 10Gb/s, and optical regeneration [7] and demultiplexing at 160Gb/s [8] in ChG waveguides. Two strategies have been identified for improving the third order nonlinear efficiency of all-optical devices: 1) increasing the waveguide nonlinear parameter, $\gamma = \omega n_2 / c A_{eff}$ (where $A_{eff}$ is the effective area of the waveguide, $c$ is the speed of light, $n_2$ is the Kerr nonlinearity, and $\omega$ is the pump frequency) and 2) using resonant structures to enhance the local field intensity.

High index materials, such as semiconductors and ChG, offer excellent optical confinement and high values of $n_2$, a powerful combination that has produced extremely high values of γ: 200,000 $W^{-1}$ $km^{-1}$ for silicon nanowires [3], and 93,400 $W^{-1}$ $km^{-1}$ in ChG nanotapers [9]. Yet silicon suffers from high nonlinear losses due to two-photon absorption (TPA) generated free carriers [10], and even if this can be ameliorated by the use of p-i-n junctions to sweep out carriers, its intrinsic nonlinear figure of merit (FOM = $n_2 / (\beta \lambda)$, where $\beta$ is the two-photon absorption coefficient and $\lambda$ the wavelength) is very low [11]. While this FOM is considerably higher for ChG [1,12,13], the development of fabrication processes for these newer materials is at a much earlier stage.

Silica glass offers many advantages for optical applications, such as extremely low (linear and nonlinear) losses and a mature fabrication technology. However, a low refractive index contrast and very small $n_2$ make it difficult to obtain a significant nonlinear response on a length scale compatible with integrated circuits. While new fabrication technologies have allowed the realization of novel structures with high values of γ, such as free-standing silica glass nanowires ($\gamma > 600$ W$^{-1}$ km$^{-1}$) [14,15], it has not been clear whether or not silica glass is capable of forming a platform for nonlinear all-optical PICs, since nonlinear optics in silica glass integrated waveguides has so far been restricted to very high peak powers [16].

The second strategy for increasing device efficiency, i.e., the use of high Q-factor resonant cavities, has been shown to be effective in enhancing FWM in both silicon [17] and GaAs [18] integrated devices. Furthermore, extremely high Q-factors (10$^7$) have been obtained in pure silica toroids and spheres [19], enabling a range of nonlinear processes using low intensity CW light, including Raman amplification [20], optical parametric oscillation [19] and very recently third-harmonic generation into the visible [21]. However, the implementation of such structures in all-optical PICs faces a number of obstacles, ranging from unconventional fabrication processes to drastic limitations in device bandwidths, preventing high bit rate operation.

In this paper, we demonstrate low power (< 7mW) continuous-wave (CW) nonlinear optics in the form of FWM based wavelength conversion in high index doped silica glass integrated ring resonators [5,22]. In particular, the success of our device is due to the combination of a high nonlinearity parameter (γ ~ 200x standard communications fiber), very low linear and nonlinear losses, and high quality fabrication processes. The high γ arises from an enhanced $n_2$ and very tightly confined (~2 μm$^2$) modes, which in turn result from the high core refractive index and core-cladding refractive index contrast (Δn = 17%), respectively. The enhancement in $n_2$ of ≈ 4.6 x silica glass is in good agreement with Millers rule [23] relating $n_2$ to n, and so similar results are, in principle, expected from other low loss, high index glasses, such as silicon oxinitride (SiON) [24], for example. Equally important, our device shows no saturation or nonlinear (multi-photon) absorption, even up to peak intensities of 25GW/cm$^2$, or 700W peak incident power, in pulsed experiments in waveguides. Further, the low loss of our device raises the possibility of cascading multiple devices on a single chip to achieve dramatically higher efficiencies. The device presented here has a bandwidth (3GHz) compatible with operation at 2.5Gb/s, and by employing higher order filter designs larger bandwidths (up to 1nm or more) [5] are achievable. All of these factors, together with the fact that this material system [5,22] has been shown to be capable of reducing PIC dimensions by up to two orders of magnitude point to low loss high index doped silica glass as a potentially attractive platform for practical all-optical photonic integrated circuits for future telecommunications systems.

Figure 1a shows a schematic of the device, a four port micro-ring resonator with radius $\cong$ 48μm, and Figure 1b shows a scanning electron microscope (SEM) picture of the waveguides cross-section (1.45 μm x 1.5 μm). Figure 1c shows the electric field modal distribution for transverse-magnetic (TM) polarized light (i.e., vertical in Figure 1b). The bus waveguides, which have the same cross section as the ring, are buried in $SiO_2$ beneath the ring and are used to couple light in and out of the resonator. The waveguide core is comprised of low loss, high index (n=1.7) doped silica glass with a core-cladding contrast of 17% [5,22], similar to other glasses such as silicon oxinitride (SiON). In addition to enhancing the nonlinear performance, the high core-cladding index contrast in this glass makes tight $\cong$ 20μm radius bends possible [22]. The unique character of our glass system [22] relates primarily to CMOS compatible fabrication: our as-deposited films are low loss (near 1550nm), in contrast with low loss SiON where very high temperature annealing is required ( > 1000C [24]). Our films were deposited by standard chemical vapor deposition (CVD) [22] and device patterning and fabrication were performed using standard photolithography and reactive ion etching to produce exceptionally low sidewall roughness on the core layer, before over-coating with a silica glass upper cladding layer. By comparing loss measurements in waveguides of different lengths on the same chip, we obtain a waveguide to fiber pigtailing loss of $\cong$ 1.5dB / facet, and waveguide propagation loss $\cong$ 0.06dB/cm, resulting in a total insertion loss (fiber to fiber) of $\cong$ 3dB. These propagation losses are almost as low as the best low-index contrast silica waveguides [27], and represent a key advantage of our waveguides compared to semiconductor and ChG devices where losses tend to be significantly higher [3].

In Figure 1d we present the INPUT-DROP transfer function from $\lambda$ = 1548.2nm to 1553.6 nm for both transverse-electric (TE) and TM polarizations, showing a Q-factor of $\cong$ 65,000, a free-spectral range (FSR) of 575 GHz, and a full-width-half-maximum (FWHM) of 3GHz.

The effect we demonstrate here (FWM) is a third order nonlinear process, where two pump photons at frequency $\omega_{Pump}$ are converted, inside a dielectric medium, into one idler and one signal photon at the frequencies $\omega_{Idler}$ and $\omega_{Signal}$ respectively, in accordance with energy conservation:

$$\omega_{Idler} = 2\omega_{Pump} - \omega_{Signal} \quad (1)$$

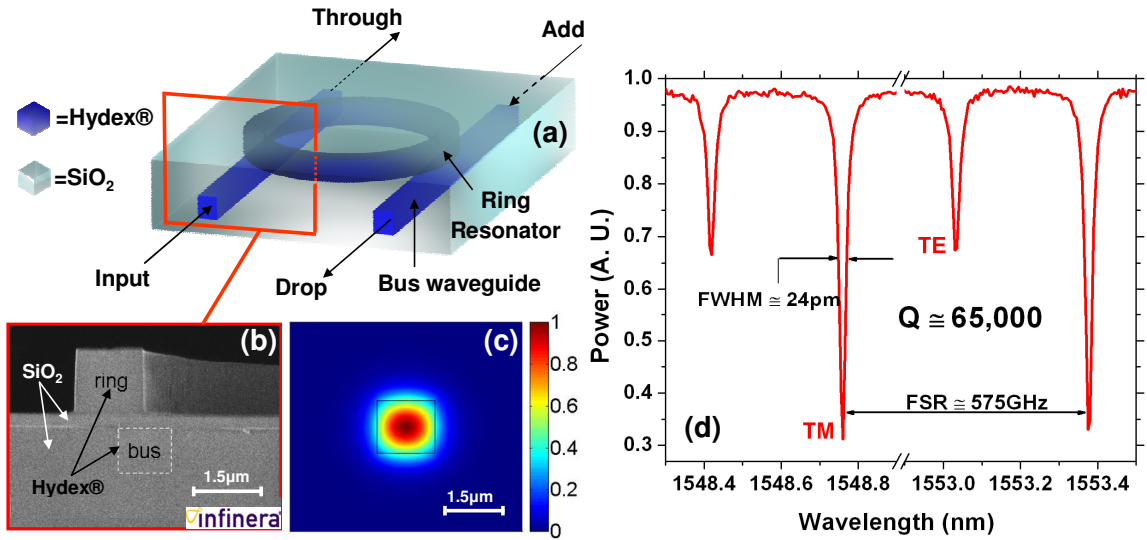

Figure 1. (a) Schematic of the 4 port micro-ring resonator (the input and output fibers of our pigtailed device are not shown). At resonance, the light injected from the INPUT port travels in the ring and exits at the DROP port, whereas the light injected in the ADD port exits the ring from the THROUGH port. During the FWM experiment, the two idlers exit from both the DROP and THROUGH ports. (b) SEM picture of the ring cross section before depositing the upper cladding of $SiO_2$. Hydex® refers to the tradename of the high index (1.7) core doped silica glass [22]. (c) Electric field modal distribution for a TM polarized beam. (d) Linear transmission through the ring resonator from the INPUT port to the THROUGH port for a TM (deeper trough) and a TE (shorter trough) polarization.

Although this process can occur with (classical) or without (spontaneous) an input signal at $\omega_{Signal}$, the classical effect is much stronger and is the basis of numerous all-optical signal processing schemes in integrated devices and optical fibers [1,28-31]. As in other nonlinear phenomena, such as second harmonic generation (SHG), phase-matching is typically required to yield appreciable efficiencies. This condition is more easily fulfilled in FWM experiments than in SHG experiments, since the involved frequencies can be very close to each other.

We performed low power experiments with a 7mW (incident power, corresponding to 5mW in the waveguide) CW pump laser tuned to a ring resonance at 1553.38nm (TM polarization) and directed to the INPUT port, while a signal laser (TM polarization) with a power of 570μW was tuned to an adjacent resonance at 1558.0 nm and directed to the ADD port. The two outputs were analyzed using either a power meter or a spectrum analyzer. Figure 2 shows the resulting output power spectra (TM polarization) as recorded from both the THROUGH and the DROP ports; experiments carried out with TE polarized modes led to similar results.

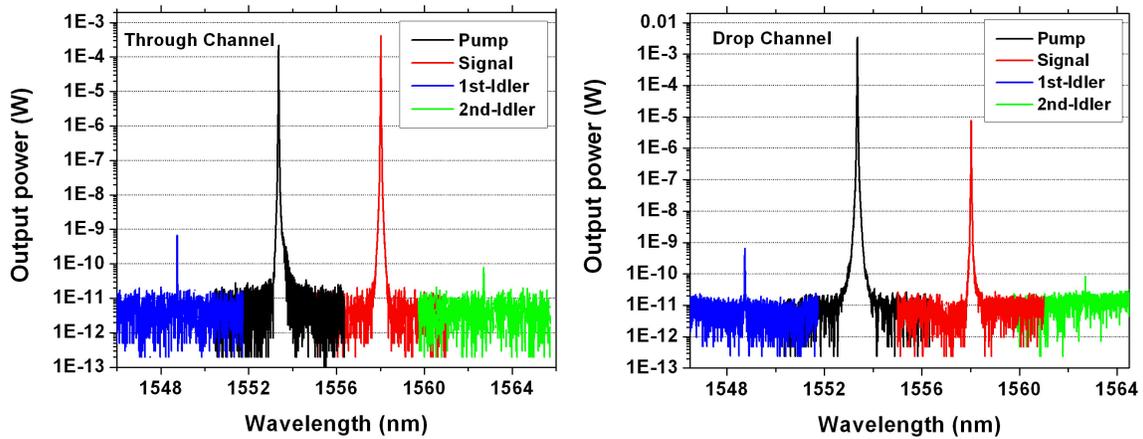

Figure 2. Experimental FWM results. Output power spectra at the THROUGH port (left) and at the DROP port (right) for TM polarized light.

Two detectable idlers were generated by the FWM process, the first according to equation (1) and the second by using the same formula but formally exchanging the role of the pump and the signal. The output power (in the bus waveguides) of the first idler was ≅ 930 pW, whereas the second idler had a power of ≅100 pW. The ratio of the output powers for the two idlers agrees remarkably well with the ratio of the pump to the signal power, $P_{idler(-1)}/P_{idler(-2)} \cong P_{pump}/P_{signal}$, as is expected for FWM.

We found that the first idler was almost exactly on resonance at 1548.7 nm, indicating that the dispersion in the system is indeed negligible. This is clearly shown by the idler detuning curve presented in Figure 3, which confirms that the idler is on resonance and consequently the FWM efficiency is maximized when the signal is on resonance [31].

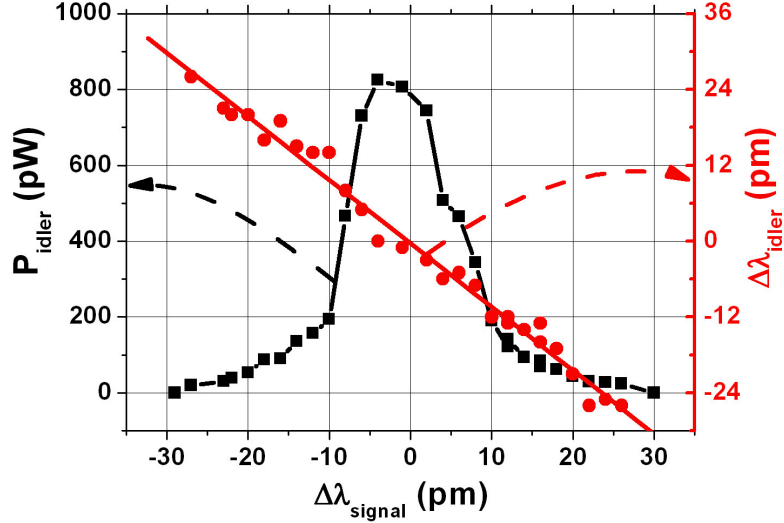

Figure 3. In black, the 1st idler power as a function of the signal detuning (from the ring resonance). In red, the 1st idler detuning as a function of the signal detuning. The linear fit passes through zero, indicating that the idler frequency from the FWM process matches the ring resonance and that the system dispersion is negligible.

We analyze the results with a theoretical model for the conversion efficiency $\eta$ that takes into account the cavity enhancement factor due to the ring geometry:

$$\eta \equiv \frac{P_{idler}}{P_{signal}} = |2\pi R \gamma|^2 P_{pump}^2 \cdot (FE_p)^4 \cdot (FE_s)^2 \cdot (FE_i)^2 \qquad (3)$$

$$FE_\mu = \frac{\sqrt{2(1-\sigma_\mu)}}{2(1-\sigma_\mu)+\alpha_\mu \pi R} \qquad \sigma_\mu = \left(1 - \frac{\pi}{2 Finesse_\mu}\right) \exp\left(\frac{\alpha_\mu \pi R}{2}\right) \qquad (4)$$

where $R$ is the ring radius, $FE_\mu$ describes the field enhancement of the ring and $\alpha_\mu$ is the mode linear loss coefficient for the mode $\mu$. Finally, $\sigma_\mu$ is the self-coupling coefficient between ring and channels. The product $(FE_p)^4(FE_s)^2(FE_i)^2$ identifies an overall field enhancement factor. Eqs. (3,4) are equivalent to those reported earlier by Absil et al. [18], although the details of the FWM considered there are different.

Figure 4 depicts the output idler power as a function of the square of the pump power (4a) and linear signal power (4b), showing good quadratic and linear dependence, respectively, as expected from theory. This demonstrates that our device exhibits no saturation due to multi-photon absorption in the ring, up to pump powers of 13dBm (incident power), limited only by the available power of our CW source. Further, in separate experiments [32] using high peak power sub-picosecond optical pulses we have confirmed that the nonlinear (multi-photon) loss in these waveguides is negligible up to extremely high peak intensities of 25 GW/cm$^2$, corresponding to >700W peak power inside the waveguides.

Figure 5 shows our experimental *external* conversion efficiency (accounting for all losses for signal, pump and idler) as a function of the incident pump power, along with theory (solid curve). Figure 5 also shows the external efficiency (conservatively estimated from their losses) from recent experiments on CW FWM in silicon-on-insulator (SOI) based ring resonators [17]. Rather surprisingly we found that at pump powers where the silicon ring resonators saturate (~ 45 - 50mW incident power or 10 - 12mW in-waveguide), the external efficiency of our device is within a couple of dB at approximately -39.5dB. Further, at higher pump powers our device efficiency is expected to increase significantly since it does not saturate (due to the absence of multi-photon absorption). At high pump powers where high bit rate all-optical signal processing via FWM in silicon nanowires has been reported [25,26] we predict our external efficiency to be ~ -24dB (at 250mW), within ~ 5dB of that obtained in [25]. Remarkably, this occurs despite the much higher nonlinearity parameter ($\gamma$) typically obtained in semiconductor waveguides [25,26]. Finally, we note that the device reported here was primarily designed for linear filter applications, leaving room for further optimization for nonlinear applications (eg., minimizing mode field area).

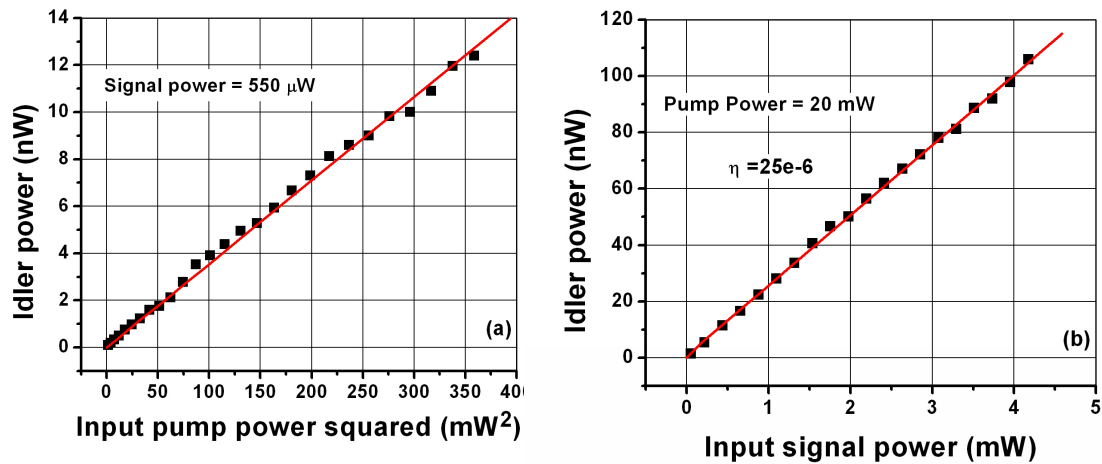

Figure 4. (a) Idler power versus the square of the pump power for FWM in the ring resonator. (b) Idler power versus the signal power.

Using the experimental data together with Eq.(3), we estimate the Kerr nonlinearity of our glass to be $n_2 = 1.15 \times 10^{-19}$ m$^2$/W, or $\approx$ 4.6 x silica glass resulting in (ogether with the small mode field size) a high nonlinearity parameter of $\gamma \cong 233$W$^{-1}$km$^{-1}$, approximately 200x standard single mode telecommunications fiber. As mentioned, since the enhancement in $n_2$ is in agreement with Miller's rule [23] we expect other high index glasses to achieve similar results as long as losses (intrinsic material as well as fabrication induced) are kept low. This is critical, since low losses were a key factor in this work, contributing to an overall field enhancement (due to the resonance) in the efficiency of $\cong 1.4 \cdot 10^7$ - orders of magnitude larger than in semiconductors where losses tend to be on the order of several dB/cm.

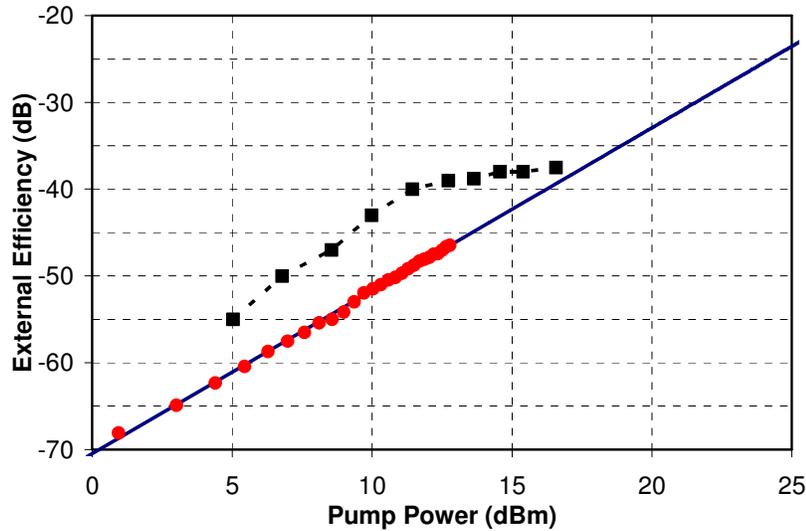

Figure 5. External efficiency versus (incident) pump power for FWM in the ring resonator: Experiment (Red points) and theory (blue line), calculated with no saturation. Also shown is the estimated external efficiency of silicon nanowire based ring resonators (50μm radius) [17] (Black data points).

In conclusion, we have presented the first demonstration of nonlinear optics in integrated, high index doped silica glass waveguides using CW light. We achieve wavelength conversion via FWM at low CW pump powers in high index doped silica glass micro-ring resonators near 1550nm, obtaining a conversion efficiency similar to (at low pump power) or dramatically higher (at high pump power) than comparable silicon devices. The low linear loss and negligible nonlinear loss of this device, together with high reliability, CMOS compatible fabrication processes, design flexibility and manufacturability, raise the prospect of high index glass as a platform for all-optical PICs for future all-optical telecommunications systems.

References

1. Eggleton, B. J., Radic, S., & Moss, D. J. *Nonlinear Optics in Communications: From Crippling Impairment to Ultrafast Tools* Ch. 20 (Academic Press, Oxford, 2008).

2. Ta'eed, V. G., et al. Ultrafast all-optical chalcogenide glass photonic circuits. *Optics*